\documentclass[%
 reprint,
 amsmath,amssymb,
 aps,
]{revtex4-2}
\usepackage{geometry}
\usepackage{graphicx}
\usepackage{dcolumn}
\usepackage{color}
\usepackage{comment}
\global\long\def\imsize{0.83\columnwidth}
 \global\long\def\halfsize{0.45\columnwidth}
\usepackage{bm}
\usepackage{comment} 
\usepackage{hyperref}
\hypersetup{colorlinks=true, citecolor=blue, urlcolor=blue, linkcolor=blue}

\begin{document}


\title{Wada Boundaries on a Hyperbolic Pair of Pants}

\author{Pedro Henrique Barboza Rossetto}%
\email{phbrossetto@gmail.com}
\affiliation{%
Department of Mathematics, University of Otago\\
9016 Dunedin - Otago, New Zealand
}%


\author{Vanessa Carvalho de Andrade}
\email{vcandrade@unb.br}

\author{Daniel Muller}
\email{dmuller@unb.br}
\affiliation{%
Instituto de F\'{\i}sica, Universidade de Bras\'{\i}lia \\
70919-970 Bras\'ilia - DF,  Brazil
}%


\date{\today}

\begin{abstract}
In this paper the geodesics of an open multiply connected hyperbolic manifold are presented from the dynamical system point of view. The approach is completely numerical. Similar to the closed hyperbolic case there is a zero-measure set of periodic orbits. The difference is that now, most of the geodesics escape to infinity through one of the topologically distinct channels, and initial conditions are identified to which channel the orbit escapes. The initial condition mesh reveals basins of attraction to each channel. We verified that the basin boundary points are of the Wada type by two independent methods. We have also calculated the basin entropy of the system, and verified that it depends on the opening angle of the escape channels. The manifold chosen is the pair of pants together with the leaves, so there are $3$ distinct exit channels.  

\end{abstract}

\maketitle


\section{\label{sec:level1} Introduction}
\textit{Wada boundaries} were first introduced by Yoneyama \cite{yoneyama1917Theory}, who considered a mathematical system which possesses the property that every boundary point is a boundary of all regions. This was first considered purely from a topological point of view, but  Kennedy and Yorke \cite{kennedy1991Basins} have reintroduced this concept to physical and dynamical systems. 

Since then, plenty of other physical systems have also been found to have the Wada property, such as chaotic scattering \cite{poon1996Wada}, hydrodynamic flow \cite{toroczkai1997Wada}, Hénon-Heiles system \cite{aguirre2001Wada}, ecological models \cite{vandermeer2004Wada}, among others. Other notable Wada systems are found in General Relativity, such as light scattering by binary black holes \cite{daza2018Wada} and geodesics of pp-wave spacetimes \cite{rossetto2020Wada}.

Numerical techniques were developed to ascertain if a boundary is of the Wada type, each one of them relies on different topological properties of Wada systems \cite{wagemakers2020How}. The main techniques are: the saddle-straddle \cite{wagemakers2020saddlestraddle}, the grid \cite{grid_wada}, and the merging \cite{merging_wada} methods.

As mentioned previously, a particular instance of the Wada property is in chaotic scattering \cite{poon1996Wada}. In some of these systems, one can identify well-defined final states and pose the question of the connection of these states with the system's initial conditions. This question is answered by graphs called \textit{escape basins}, where initial conditions are displayed in a two-dimensional plane, and the final estate is represented by a colour \cite{kennedy1991Basins}.

A very fundamental question in astrophysics is what is the topology of the Universe \cite{Luminet:2003dx}. This question can only be measured or answered from physical effects that result from the Universe being multiply connected. 

It is well-known topology is not completely defined by Einstein's General Relativity. For instance the flat torus $\mathbb{T}^2$ or the euclidean plane $\mathbb{E}^2$ are indistinguishable according to GR. Non-trivial topology is performed by identifications of sides or faces by elements of a discrete subgroup $\Gamma$, of the full group of isometries \cite{dubrovin2012modern}. If this subgroup acts freely and properly discontinuous the resulting manifold satisfies the Hausdorff property \cite{lee2013introduction} \cite{lee2010introduction}, which states that two disjoint points always have disjoint neighbourhoods.

One of the most direct consequences of non-triviality of space is the occurrence of multiple images. There is always a direct orbit between a distant galaxy and the observer, a geodesic. While the indirect geodesics go through one of the identified sides, re-enter the region and reach the observer. This process can repeat itself several times, the geodesic re-entering the region each time and in the end reaching the observer.

The first to discuss the possibility of the occurrence of multiple images for non-trivial topology was Karl Schwarzschild in 1900 considering a sufficiently big (old) Universe, for a translation see \cite{1998}. Pioneering work of cosmic topology started with Ellis \cite{Ellis:1970ey}, \cite{sokolov1974estimate}, Fagundes \cite{fagundes1985quasar} and \cite{fang_sato1985}. 
There is an excellent review on cosmic topology by Lachi\`eze-Rey and Luminet \cite{LACHIEZEREY1995135} after which there was an increase in interest in this subject especially connected to the possibility of measurements on CMBR. Other reviews are for example \cite{levin2002topology} and in this next reference there are several articles resulting from Cleveland Workshop on Cosmic Topology held at Case Western Reserve University in 1997 \cite{CWCT_98,Starkman_1998}. See also \cite{Cornish:2003db,Gomero:1998dz,Gomero:1999ut,Aurich:2013fwa,Vaudrevange:2012da,Luminet:2003dx}.

In physics, the importance of geodesic structure is its relation to Green functions, for example in Feynman's summation over classical paths approach, which can also be used to obtain vacuum expectation values \cite{Camporesi:1990wm,DeWitt:1979dd}. These summations can be both continuous or discrete \cite{terras2012harmonic, terras2013harmonic, helgason1981topics, vilenkin1969fonctions}, and in harmonic analysis they are also known as the method of images which of course is the same as principle of the multiple images mentioned above applied to functions \cite{hejhal1976selberg}.

It is very well known that geodesics on closed hyperbolic manifolds are among the highest chaotic systems known  \cite{anosov1969geodesic, Balazs:1986uj}. 
On the other hand, apparently this topic has not been addressed for open manifolds, which is the subject of this present work. 

In this article we choose the pair of pants manifold which is in fact the pair together with the leaves, and we decided to call the $\mathbb{Y}$ manifold, which is modelled in the hyperbolic plane $\mathbb{Y}=\mathbb{D}/\Gamma$. This manifold naturally presents $3$ topologically distinct exit channels. In this case the initial condition for each geodesic is identified with respect to which exit channel the geodesic reaches infinity. From the cosmological motivation point of view, the initial condition mesh reveals the basin structure of the system, the set of geodesics which can be seen from each exit channel. Of course there are $3$ basins in this system, and we verified that their boundary satisfies the Wada property by two different numerical methods. We also calculated the basin entropy to check that entropy increases as the angle of aperture to infinity decreases.

We divide the present paper as follows: Section \ref{sec2} reviews very well-known results in hyperbolic geometry in the Poincar\'e disc model; Section \ref{sec3} discusses the pair of pants topology and tilling of the hyperbolic space; Section \ref{sec4} discusses the dynamical aspects of geodesics in the hyperbolic pair of pants; Section \ref{sec:wada} proves that the system posses the Wada property by two different numerical methods; and, finally, Section \ref{sec:basin_entropy} establishes the relation between the exit angle and the basin entropy of the system. 

\section{The Poincar\'e disc}
\label{sec2}
This section intends to give a brief overview of the geodesics of the hyperbolic plane $\mathbb{H}^2$. For further details, see for example the excellent review article by Balazs and Voros \cite{Balazs:1986uj}.

For this we start with disc model of $\mathbb{D}$ for this geometry 
\begin{align}
ds^2=\frac{4(dx^2+dy^2)}{(1-x^2-y^2)^2},
\end{align}
with $r=\sqrt{x^2+y^2}<1$. 
It is very well known that in $\mathbb{D}$ the group of isometries are the elements $g\in SU(1,1)$ 
\begin{align}
&g=\left( \begin{tabular}{cc}
   $\alpha $ &  $\beta$ \\
    $\beta^*$ &$\alpha^*$ 
\end{tabular}\right) & \mbox{with  } |\alpha|^2-|\beta|^2=1
\label{elem.g}
\end{align}
and the action of the group is non-linear and given by homographies or M\"obius transformations 
\begin{align}
z\rightarrow \frac{\alpha z +\beta}{\beta^*z+\alpha^*},
\label{homografias}
\end{align}
with $z=x+iy$. 

Also, as is well known, considering the Lagrangian 
\begin{align}
    & \mathcal{L}=\frac{2(\dot{x}^2+\dot{y}^2)}{(1-r^2)^2} & H=\frac{2(\dot{x}^2+\dot{y}^2)}{(1-r^2)^2}
   \label{L-H} 
\end{align}
with corresponding Hamiltonian $H=\mathcal{L}$ \footnote{Energy and phase space is used in a loose sense, strictly speaking $H$ is just a conserved quantity. We also leave the velocities $\dot{x}$ and $\dot{y}$ in the Hamiltonian.}  which is conserved and according to Noether's theorem, the following conserved quantities 
\begin{align}
&L=\frac{4(x\dot{y}-y\dot{x})}{(1-r^2)^2}, & B_1=\frac{2\dot{y}}{1-r^2}+xL\nonumber\\
& B_2=-\frac{2\dot{x}}{1-r^2}+yL
\label{p_conservado}
\end{align}
are obtained, and it is an algebraic exercise to evaluate $\vec{B}.\vec{B}=B^2=(B_1)^2+(B_2)^2$
\begin{align*}
B^2=2H+L^2. 
\end{align*}

The above equation has two consequences, the first results into the geodesic orbit as follows. Considering the vector $\vec{r}=(x,y)$ with $x$ and $y$ in $\mathbb{D}$ take the scalar product $\vec{r}.\vec{B}=(1-r^2)L/2+r^2L$ so that it is easy to obtain 
\begin{align}
   (\vec{r}-\vec{B}/L)^2=(B/L)^2-1=\frac{2H}{L^2}
   \label{geodesica}
\end{align}
 which shows that the geodesics in $\mathbb{D}$ are circles centered at $\vec{B}/L=(X_c,Y_c)$ with $|\vec{B}/L|>1$ and radius $\sqrt{2H}/|L|$. These circles are orthogonal to the circle $r=1$ at infinity with center that lies outside $\mathbb{D}$. In complex notation $z=x+iy$ with $x,\; y\subset \mathbb{D}$ and $|z|\leqslant 1 $ the geodesic is written as 
 \begin{align}
     |z-s|=\sqrt{|s|^2-1},
     \label{g_complexa}
 \end{align}
 where $s=X_c+iY_c,$ with $|s|>1$.
The second consequence is that the Hamiltonian can be written as the combination of two conserved quantities 
\begin{align}
  &  H=\frac{1}{2}\left( -L^2+B^2\right), 
    \label{H}
\end{align}
and from which it is evident that the elements of the group $SU(1,1)$ acting on $\mathbb{D}$, correspond to elements of $SO(1,2)$ acting on the vector space $P=(L,B_1,B_2)$. The mapping of the elements is not so evident and we quote this in Appendix \ref{A1}. It is also clear that $P$ is space-like as $B^2$ must always exceed $L^2$ such that isometries acting on $P$ can vanish or also reverse the sign of $L$. 

\section{The pair of pants manifold}
\label{sec3}
One of the techniques to obtain the Fuchsian group is through Poincaré's polygon theorem. Strictly, this method was meant to be used only for closed polygons \cite{de1971polygones}, polyhedrons \cite{epstein1994exposition}. Loosely speaking, the theorem relies on the tilling of the hyperbolic plane $\mathbb{D}$ by an appropriate even-sided polygon which must not leave any blank space nor superpose any part of the moved polygons. The motions of the polygons are performed by specific elements of the group of isometries and the summation of the angles at each vertex of union of the polygons must add to $2\pi$. Poincar\'e theorem results in closed, boundaryless manifolds which are topologically equivalent to identifications of sides of generalised polyhedra. 

In this present work the focus is on open manifolds instead, and the terminology of geodesics half-planes is borrowed from the excellent book \cite{bolte2012hyperbolic} specifically in the chapter written by Aigon-Dupuy, Buser and Semmler. There must be an even number of non-intersecting sets of half-planes and $\mathbb{D}$ is covered by a tilling of such set of half-planes in Poincar\'e's sense. Since we want to consider manifolds, the set of motions that perform the tilling must only fix the $2$ points in the boundary $\partial \mathbb{D}$, the action is called free and properly discontinuous. The group which satisfies these properties is called Fuchsian and is isomorphic to the fundamental group $\Gamma$ of the manifold. 

\begin{figure}[htpb]
 \begin{center}
\begin{tabular}{c }
\resizebox{\imsize}{!}{\includegraphics{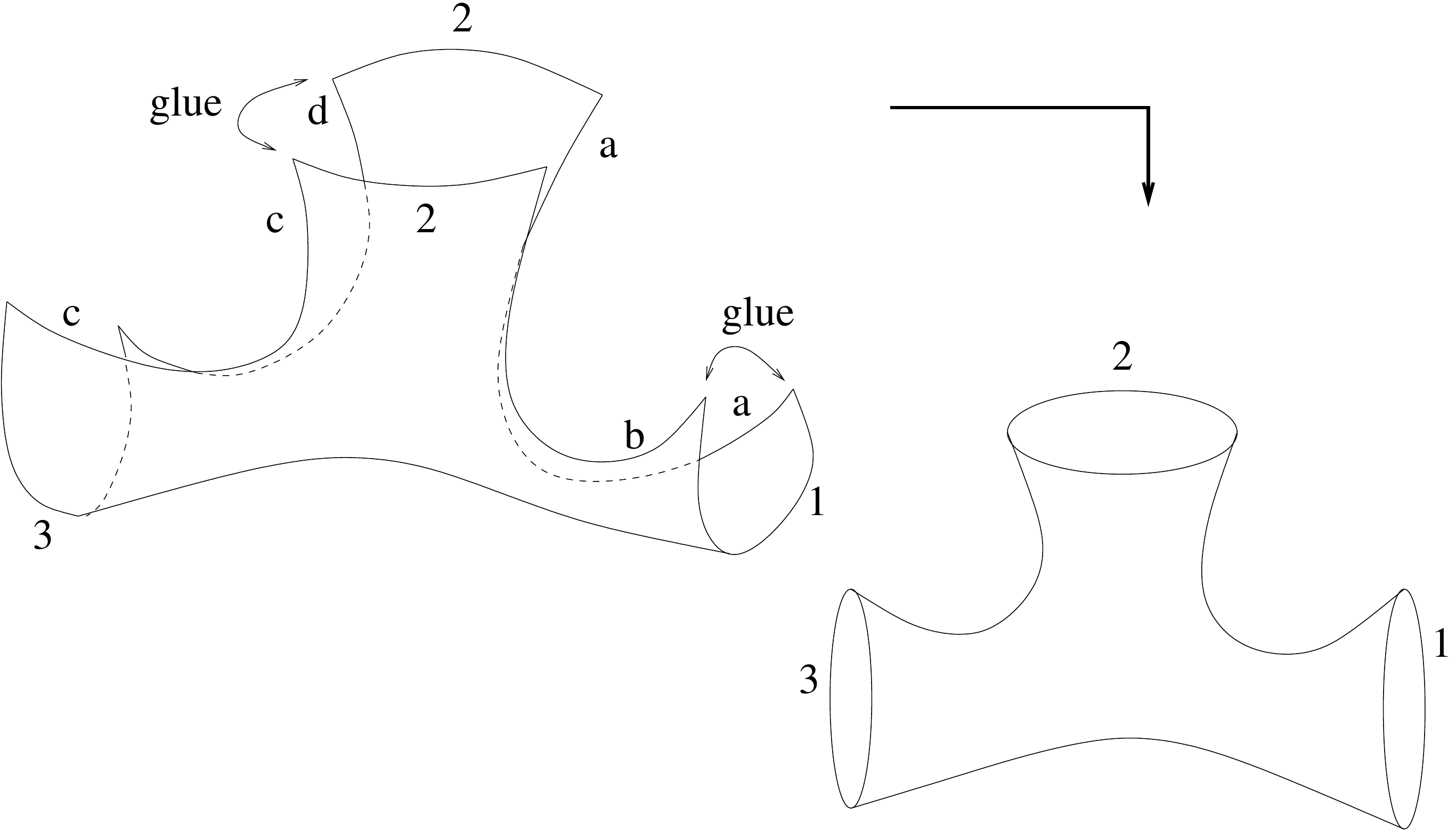}} 
   \end{tabular}
    \end{center}
    \caption{This figure shows the $\mathbb{Y}=\mathbb{D}/\Gamma$  manifold through the side paring performed by the generataros $\mbox{g}_1$ and $\mbox{g}_2$ in \eqref{trans_s_1} and \eqref{trans_s_2} respectivelly. \label{fig3}}
    \end{figure}

The set of half-planes is the fundamental region $\mathcal{F}$ of the manifold which is obtained by identification of appropriate sides of $\mathcal{F}$ as shown in FIG. \ref{fig3}, and we decided to call $\mathbb{Y}=\mathbb{D}/\Gamma$ the resulting manifold. 
The side pairings are performed by specific elements of the group of isometries which are the generators of the fundamental group $\Gamma$. For $\mathbb{Y}$ there are only two generators which we call $\mbox{g}_1$ and $\mbox{g}_2$.

In the following we will obtain these generators and the fundamental group $\Gamma$ for this manifold. Again, \cite{bolte2012hyperbolic} explains this in a didactic way. For this reason, it is needed a set of non-intersecting geodesics. First consider the geodesic written in \eqref{g_complexa} $|z-s|=\sqrt{|s|^2-1}$. Then, the following group element 
\begin{align}
  \mbox{g}_1= \frac{1}{\sqrt{|s|^2-1}}\left( \begin{tabular}{cc}
       $s$&$1$  \\
        $1$ &$\bar{s}$
    \end{tabular}\right)
    \label{trans_s_1}
    \end{align}
sends geodesic  
\begin{align}
|z-s|=\sqrt{|s|^2-1}    
\label{g.1}
\end{align}
to geodesic
\begin{align}
|z+\bar{s}|=\sqrt{|s|^2-1},
\label{g.2}
\end{align}
that is, performs the reflection of this specific geodesic \eqref{g_complexa} with respect to the $y$ axis. Also following \cite{bolte2012hyperbolic}, the geodesic 
\begin{align}
&|z-q|=\sqrt{|q|^2-1} & q=\frac{-2}{s-\bar{s}}
\label{g_central1}
\end{align}
is named the central geodesics which is sent to itself by \eqref{trans_s_1}. It is not difficult to see that a generic element of \eqref{elem.g} can be written as 
\begin{align}
  \mbox{g}_2= \frac{1}{\sqrt{|u|^2-1}}\left( \begin{tabular}{cc}
       $\bar{u}$&$-e^{i\beta}$  \\
        $-e^{-i\beta}$ &$u$
    \end{tabular}\right)
    \label{trans_s_2}
    \end{align}
which sends geodesic 
\begin{align}
    |z+\bar{u}e^{i\beta}|=\sqrt{|u|^2-1}
    \label{g.3}
\end{align}
to geodesic 
\begin{align}
    |z-ue^{i\beta}|=\sqrt{|u|^2-1}.
    \label{g.4}
\end{align}
The central geodesic in this case is 
\begin{align}
& |z-v|=\sqrt{|v|^2-1} & v=\frac{-2e^{i\beta}}{u-\bar{u}}.
\label{g_central2}
\end{align}

\begin{figure}[htpb]
 \begin{center}
\begin{tabular}{c c}
\resizebox{\halfsize}{!}{\includegraphics{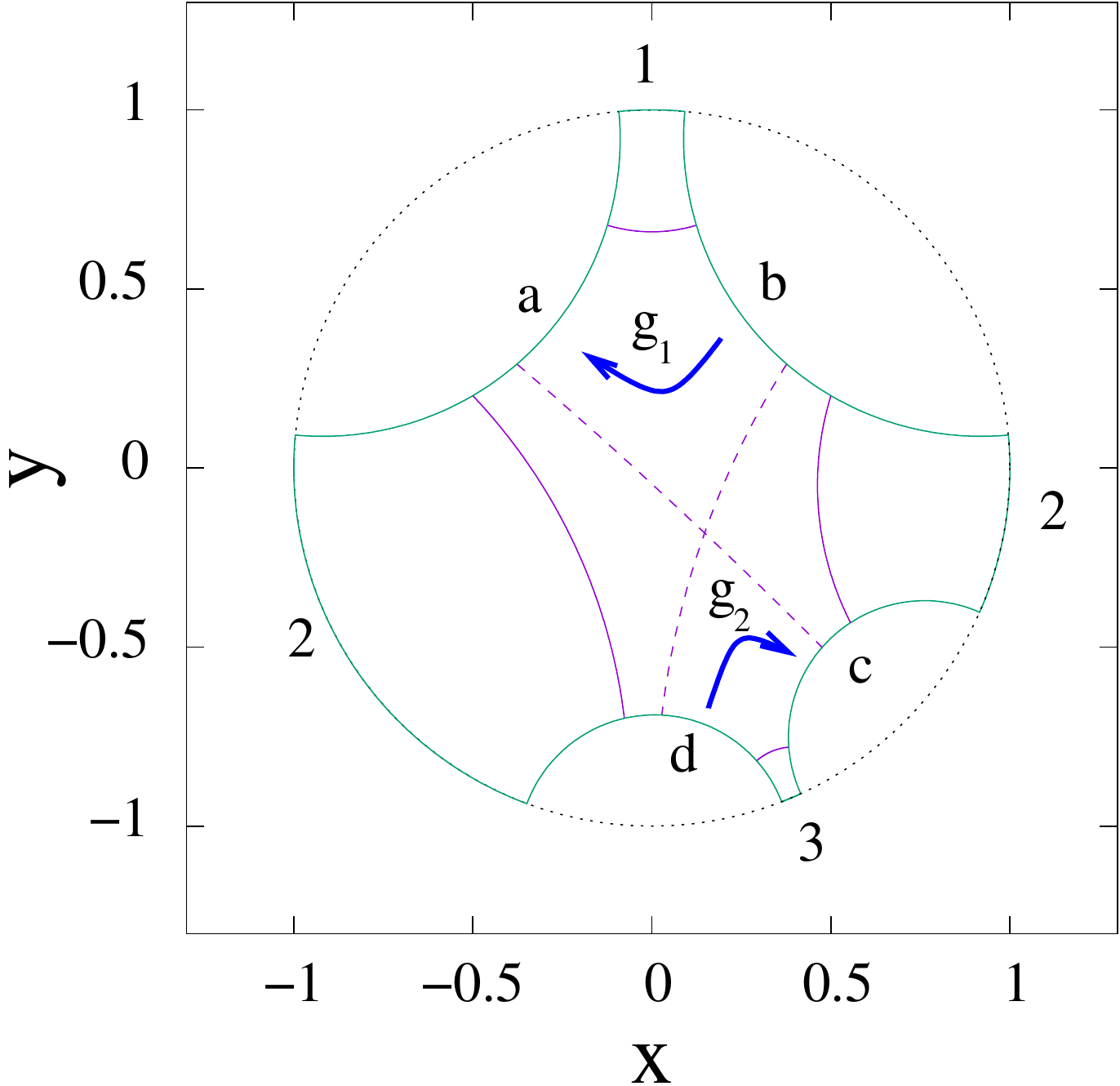}} &  
       \resizebox{\halfsize}{!}{\includegraphics{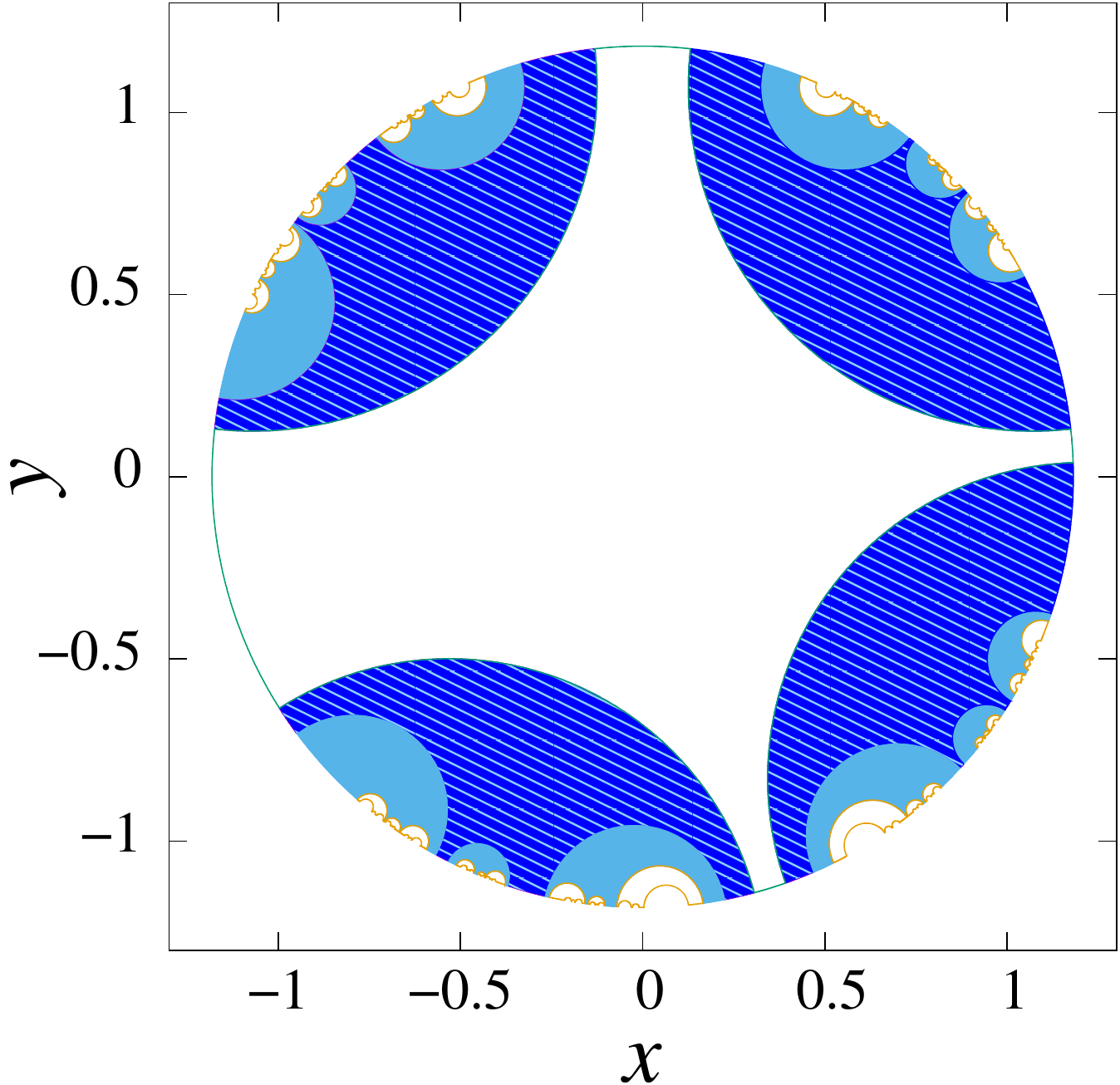}}  \\
       a) & b) 
   \end{tabular}
    \end{center}
    \caption{Panel a) Fundamental region $\mathcal{F}$ with the respective boundaries at infinity numbered $1,\,2,\,3$. The elements $\mbox{g}_1$ and $\mbox{g}_2$ that perform the identification of the sides of $\mathcal{F}$ are also shown $b \rightarrow a$ and $d \rightarrow c$. For this example manifold, $s=1.3e^{i\pi/4}$, $u=1.07e^{-i3\pi/8}$ and $\beta=0.4$. It is also shown in dashed a periodic orbit and the solid lines that separate the funnels that are also periodic orbits. Panel b) Tessellation of $\mathbb{D}$ with copies of the fundamental polytope according to the motions sketched in panel a). \label{fig1}}
    \end{figure}

The set of $4$ geodesics, \eqref{g.1}, \eqref{g.2}, \eqref{g.3} and \eqref{g.4} form the fundamental region $\mathcal{F}$ of the manifold which is an infinite, non-closed polytope shown in FIG. \ref{fig1}a). Then, $\mbox{g}_1$ and $\mbox{g}_2$ given respectively by \eqref{trans_s_1} and \eqref{trans_s_2} are the generators of the fundamental group $\Gamma$. The elements, $\gamma_i\in \Gamma$ are obtained by all possible non-equivalent products of $\mbox{g}_1$ $\mbox{g}_2$ and their inverses. The action of the elements of $\Gamma$ move the fundamental polytope forming a tilling of hyperbolic space which covers the entire disc $\mathbb{D}$ as shown in FIG. \ref{fig1}b). 
 
 Both central geodesics,\eqref{g_central1} and \eqref{g_central2} are also shown in FIG. \ref{fig1}a). These two geodesics are periodic and define the upper and lower funnels respectively numbered $1$ and $3$ in FIG. \ref{fig1}a). They are periodic of period 1, since they are invariant under the action of $\mbox{g}_1$ or $\mbox{g}_2$ respectively. 
 
 The funnels numbered $2$ and $3$ are also defined by a pair of periodic geodesic and are shown in FIG. \ref{fig1}a). This pair is obtained as the central geodesic of the following products $\mbox{g}_1\mbox{g}_2$ and $\mbox{g}_2\mbox{g}_1$, respectively. These geodesics are periodic of period two since it is necessary the action of two generators to bring the geodesic back to itself. 
 
 FIG. \ref{fig1}a) also shows another pair of periodic geodesic in dashed which are the central geodesics of the products $\mbox{g}_1(\mbox{g}_2)^{-1}$ and $(\mbox{g}_2)^{-1}\mbox{g}_1$. This case is also period 2. 
 
 The non-equivalent products of the generators and their inverses are the elements of the fundamental group $\Gamma$. In a non-rigorous sense, the elements of the fundamental group can be mapped to the set of integers and each one of these elements has its respective periodic geodesic. Thus the periodic geodesics have zero measure in the set of geodesics.

\section{Escape Basins on the $\mathbb{Y}$ manifold}
\label{sec4}
The geodesics in $\mathbb{Y}$ naturally reduce to an exit system with $3$ topologically distinct possibilities, and it is necessary to choose a constant energy surface, H given by \eqref{L-H} or \eqref{H} which is what is done in the following. In short, exit systems are obtained by cutting holes in phase space and check what set of initial conditions hit these holes, for a standard reference on exit systems, see for example Ott's excellent book \cite{ott2002chaos}. 

We decided to start all geodesics, $|z-c|=\sqrt{|c|^2-1}$ with $c=|c|e^{i\alpha}$ and $|c|>1$ from the point of closest approach to the origin of $\mathbb{Y}$, see FIG. \ref{fig1}a), $z_i=x_i+iy_i=e^{i\alpha}(|c|-\sqrt{|c|^2-1})$  and positive initial $L$, $L>0$ \eqref{p_conservado}. That is, all initial conditions $z_i\rightarrow (x_i,y_i)=(|c|-\sqrt{|c|^2-1})(\cos(\alpha),\sin(\alpha))$ for the geodesics, are points which are inside  $z_i \subset \mathcal{F}$ that move initially with zero radial velocity and increasing polar angle with respect to the origin of $\mathcal{F}$. Besides that, we also choose $H=1/2$ in \eqref{H} which does not introduce additional restrictions. The geodesic is then completely specified by the complex number $c=|c|e^{i\alpha}$, the conserved quantities in \eqref{p_conservado} can be recovered as $B/|L|=|c|$  and $\vec{B}=B(\cos(\alpha),\sin(\alpha))$. Remind that as mentioned above, the vector $\vec{B}/L$ defines the radius vector of the origin of the geodesic according to \eqref{geodesica}.

\begin{figure}[htpb]
 \begin{center}
\begin{tabular}{c c}
\resizebox{\halfsize}{!}{\includegraphics{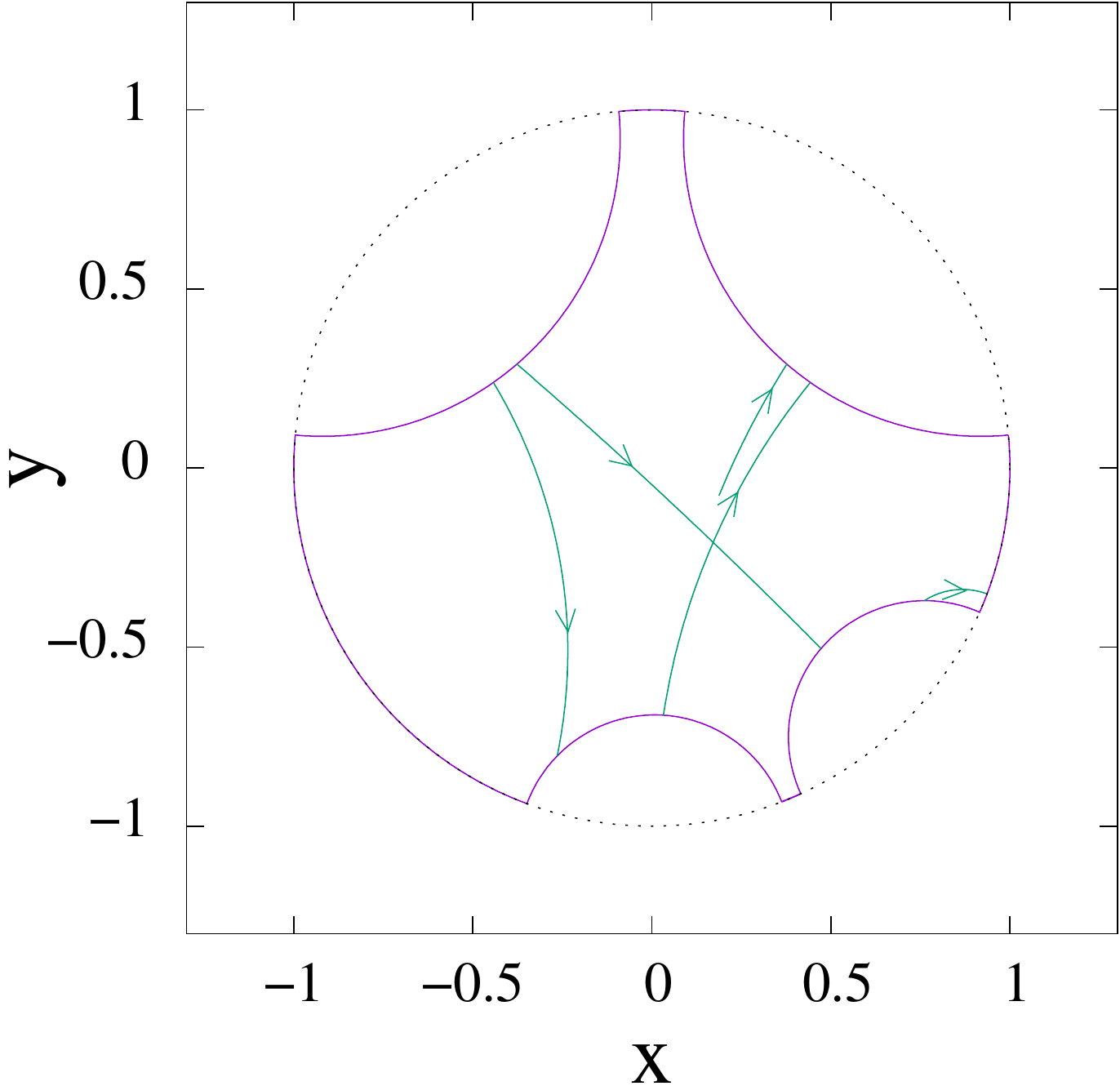}}  &
\resizebox{\halfsize}{!}{\includegraphics{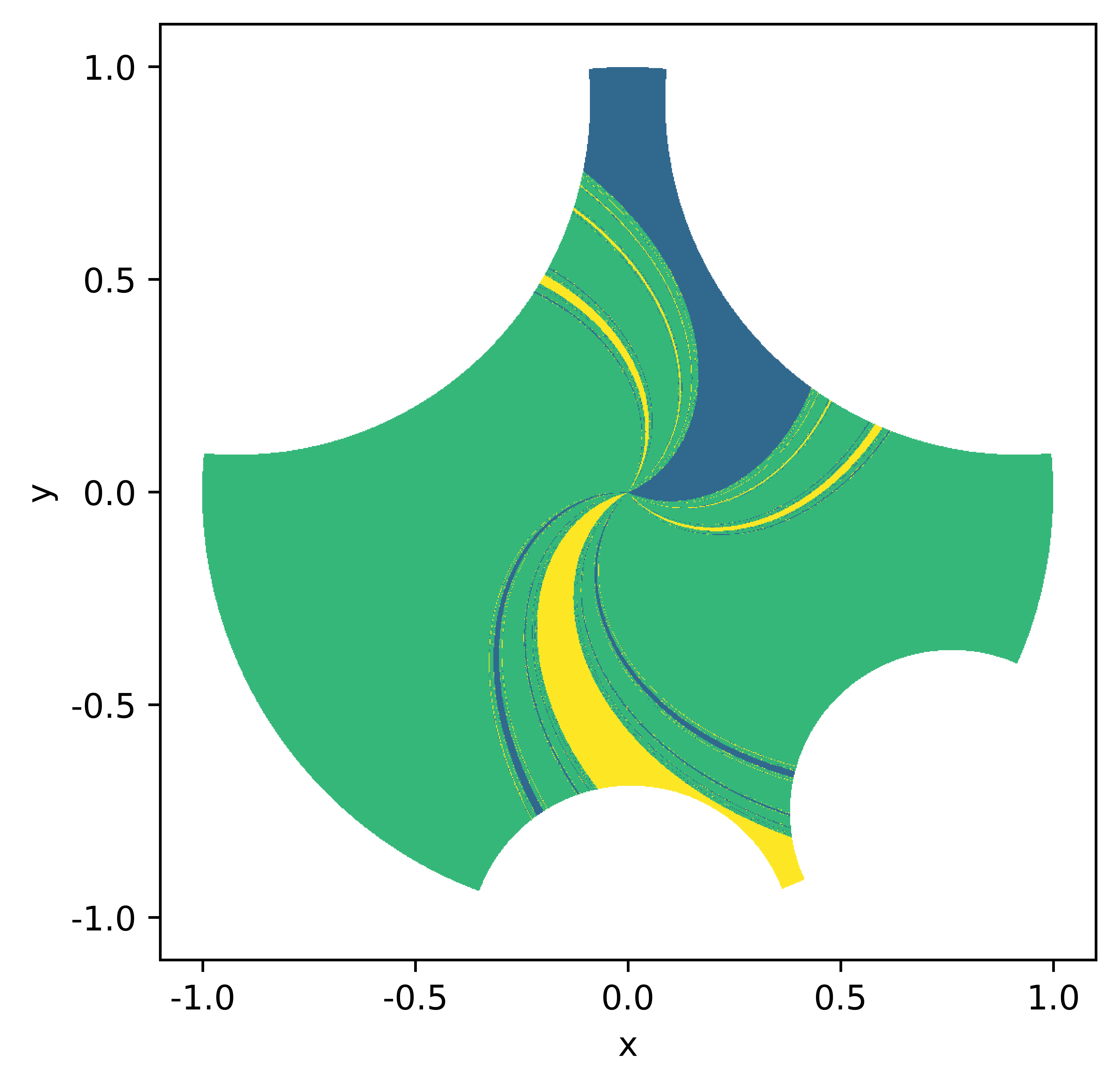}}   \\ 
       a) & b) 
   \end{tabular}
    \end{center}
    \caption{a) This plot shows a geodesic with initial conditions $|c|=2.568$ and $\alpha=-0.389\mbox{rad}$ which hits exit $2$. b) Basin plot for the surface $H=1/2$ as described above. Black points are outside $\mathcal{F}$, dark green points to exit 1, light green points to exit 2, and yellow points to exit 3. The manifold is specified in FIG. \ref{fig1}a).  \label{fig4}}
    \end{figure}

This arbitrary geodesic will hit either one of the sides of  $\mathcal{F}$, $a$, $b$, $c$, $d$ or one of the boundary intervals at infinity, named $1$, $2$ and $3$ as shown in FIG. \ref{fig4}a).
Since $\mathbb{Y}$ is non-trivial, a single geodesic can wind around a few times and almost always go to infinity, reaching either exit $1$, $2$ or $3$. In FIG. \ref{fig4}a) a geodesic is shown which reaches exit $2$. 

As mentioned at the end of \ref{sec3}, although most of the geodesics are non-periodic, there is a zero-measure set of periodic orbits which do not hit any exit, as shown in FIG. \ref{fig1}a). 

Since most of the geodesics in $\mathbb{Y}$ escape through the topologically different exits 1, 2 and 3, it is natural to ask what set of initial conditions are attracted to a particular exit. We can answer this question by analysing the \textit{escape basin}.

The escape basin is the association of each pair of initial conditions with the number of the exit that the respective geodesic took to escape to infinity. We can display this by making a graph where the initial conditions are set in a pixel grid. The colour of each pixel is associated with one of the exits.

In this spirit, in FIG. \ref{fig4}b) it is also shown a 1 million points basin plot for the specific manifold described in FIG. \ref{fig1}a). Dark green points go to exit 1, light green points to exit 2, and yellow points to exit 3.

\section{The Wada Property}
\label{sec:wada}
\begin{figure}[htpb]
 \begin{center}
\begin{tabular}{c c}
\resizebox{\halfsize}{!}{\includegraphics{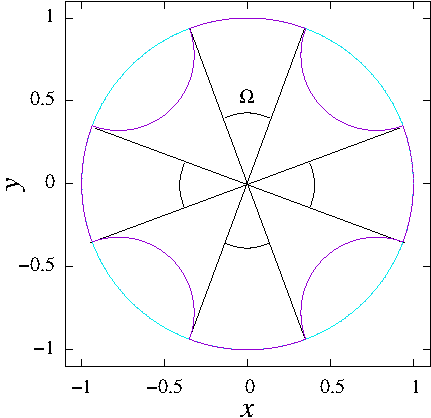}}  &
       \resizebox{\halfsize}{!}{\includegraphics{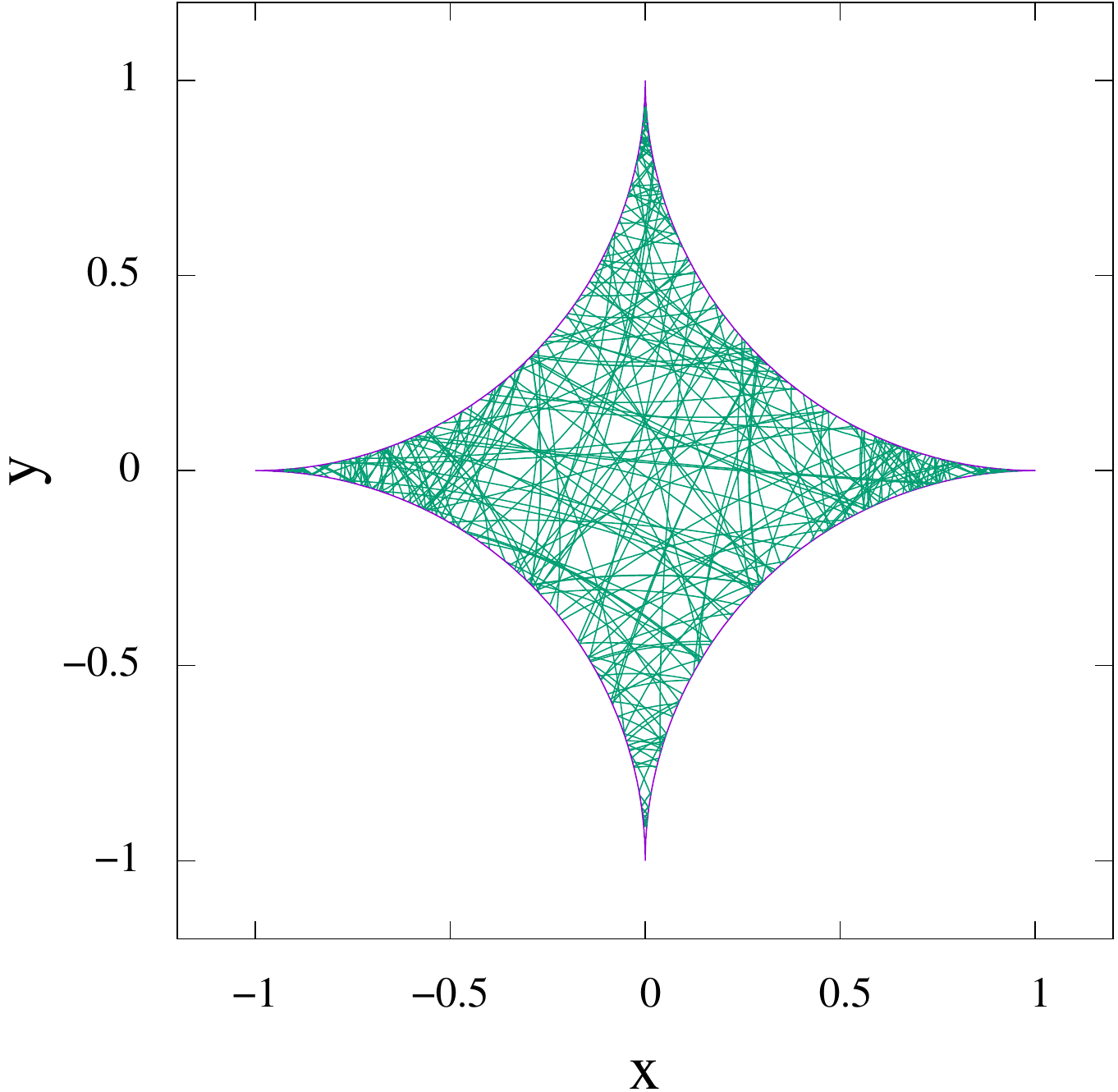}}  \\
        a) & b)
   \end{tabular}
    \end{center}
    \caption{Panel a) shows the symmetrical situation for which $u=|s|e^{-i\pi/4}$, $s=|s|e^{i\pi/4}$ and $\beta=0$ in 
\eqref{trans_s_1} and \eqref{trans_s_2}. The angle of aperture $\Omega$ is given by \eqref{aperture} and is the same for all exits. In panel b) the single parameter $s\rightarrow \sqrt{2}$ for which $\Omega\rightarrow 0$ and it is shown a single geodesic which winds around 1000 times in $\mathcal{F}$. \label{size}}
    \end{figure}

In the following a particular class of fundamental regions $\mathcal{F}$ is chosen. First consider $\mbox{g}_1$ in \eqref{trans_s_1} with $s=|s|e^{i\pi/4}$, we call the angle of aperture $\Omega$ 
\begin{align}
    &\Omega=\frac{\pi}{2}-2\arccos(1/|s|), \label{aperture}
\end{align}
as shown in FIG. \ref{size}a). Then consider $\mbox{g}_2$ given by \eqref{trans_s_2} also with $u=|s|e^{-i\pi/4}$ and $\beta=0$. This choice of group generators, $\mbox{g}_1$ and $\mbox{g}_2$ forms a class of $\mathcal{F}$ as long as $s=]1,\sqrt{2}]$, which we call symmetrical and is the one for which all aperture angles are the same, also as shown in Fig. \ref{size}.

The reason for this class of $\mathcal{F}$ is that it is easy to perform the transition to a closed hyperbolic manifold of negative curvature in the limit when $|s|\rightarrow \sqrt{2}$ as shown in FIG. \ref{size}b). As already mentioned in the introduction it is very well-known that geodesic motion in closed manifolds of negative curvature is highly chaotic \cite{anosov1969geodesic}, as it is shown in FIG. \ref{size}b).

\begin{figure*}
    \centering
    \includegraphics[height=0.5\textheight]{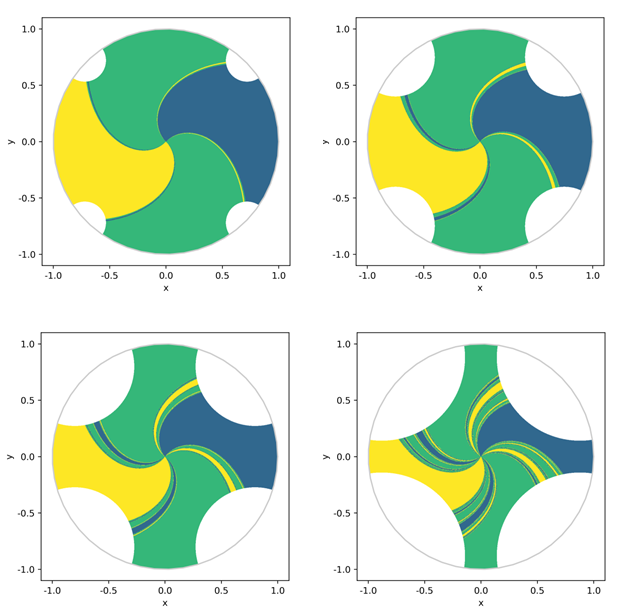}
    \caption{Some characteristics escape basins of the manifold. From top to bottom, left to right, we have $\Omega$ respectivelly equal to $1.2$, $0.9$, $0.6$ and $0.3$.}
    \label{fig_basins}
\end{figure*}

In FIG. \ref{fig_basins} we have chosen this symmetrical configuration of the geodesics that form the fundamental region $\mathcal{F}$ and we have varied the exit angle $\Omega=1.2,\,0.9,\,0.6,\,0.3$ of each exit.

It is apparent from the basins displayed that the boundaries have an intricate structure. In the following, we are going to check if the boundaries have the Wada property, which in short means that a boundary point of two regions is a boundary point of all regions.

In order to verify this, we will use the merging \cite{merging_wada} and grid \cite{grid_wada} methods.

\subsection{The Merging Method}

In the following the merging technique developed by Daza, Wagemakers and Sanju\'an \cite{merging_wada} is applied. The method acts directly to the basin points of a specific basin plot and consists in 2 steps, 1) First choose one of the colours, and then equate all of the other colours obtaining a two coloured basin. The thin boundary is formed by setting pixels with same neighbouring colour as white, say, while the other pixels coloured black otherwise, which are the boundary pixels. The set of non-equivalent thin boundary basins resulting from this first step is named $S_0$ with number of elements the same number of colours of the initial basin since each time one different colour is chosen.
2) The second step involves the successive enlargement of the thin boundary set $S_0$ and is repeated several stages as explained in the following. At the first stage named $r=1$, each neighboring pixel of a boundary pixel is painted black in $S_0$ resulting in a new set of thick boundary basins $S_1$. Then the number of boundary pixels in the set $S_0$ which are not contained in all the enlarged boundaries in $S_1$, are recorded and divided by the total number of boundary pixels in $S_0$. The previous stage is repeated with $r=2$ starting from $S_1$ to $S_2$ recording all boundary pixels in $S_0$ not contained in $S_2$, and dividing by the total number of boundary pixels in $S_0$ and the process is repeated for increasing $r$ until all the boundary pixels in $S_0$ are contained in the enlarged boundaries $S_r$.

\begin{figure*}
    \centering
    \includegraphics[height=0.4\textheight]{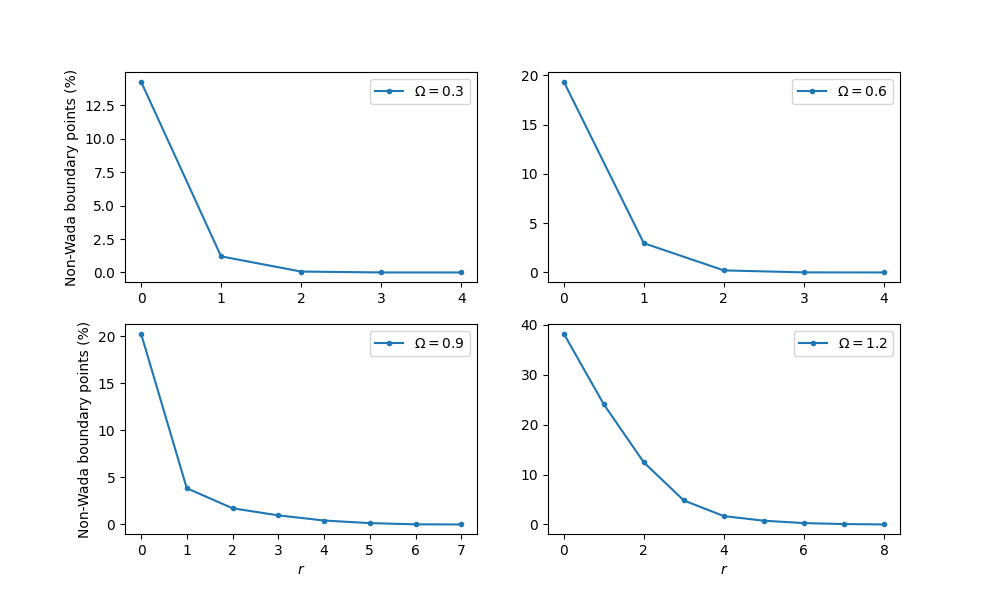}
    \caption{This plot shows the merging method at each enlarging stage. The $x$ axis is the number of enlargements of the basin boundary points $r$, and the $y$ axis is the percentage of points of the thin boundaries that are not included in the enlarged boundaries. Each one of the plots in this figure is the merging method applied to the respective basin plot in FIG. \ref{fig_basins}}
    \label{merging}
\end{figure*}

In FIG. \ref{merging}, the above procedure is applied to each basin plot shown in FIG. \ref{fig_basins}. 

\subsection{The Grid Method}

Daza et al. developed the grid method \cite{grid_wada}, and one can find the complete algorithm in their paper. In short, the method divides the escape basins into grid cells and uses the colour of neighbouring cells to classify the central one.

In our case, we have only three possible exits. Therefore, the classification of the points happens as follows: if a particular grid cell is surrounded by other cells of the same colour, the point is a \textit{interior point}; if there are two different colours, it is a \textit{simple boundary point}; and, if all three colours are present, it is a \textit{Wada point}.

At times, Wada points can be wrongly identified as simple boundaries due to problems with the basin's resolution. Because of this, each time a point is classified as a simple boundary, we make a grid refinement to check if we do not actually have all three colours present. These refinements happen $q_{max}$ times until a stability condition is met.

An important quantity for this method is the ratio between the number of simple boundary and Wada points. We define

\begin{equation}
    \label{W_def}
    W_{m}^q=\frac{N_m^{q}}{N_2^{q}+N_3^{q}}, 
\end{equation}

\noindent where $m=\{2,3\}$, $N_2^{q}$ and $N_3^{q}$ are respectively the number of simple boundary and of Wada points at iteration $q$ of the refinement process.

Finally, a system is Wada if $W_3\to1$, and consequentially $W_2\to0$, as $q\to\infty$. As this is a numerical method, $q$ does not go to infinity but to $q_{max}$ instead.

We have used this method to analyse the basins shown in FIG. $\ref{fig_basins}$, and we display the results obtained in FIG. \ref{fig_Ws}. Notice that the stability condition is that increasing the iterations $q$ does not change the values of $W$.

The important implication of the result shown in FIG. \ref{fig_Ws} is that for every basin the value of $W_3$ went to one as the number of iterations grew. And as we discussed previously, this means the boundary of the system is Wada. This corroborates the result obtained by the merging method.

\section{Basin Entropy}
Another method developed to characterise the degree of uncertainty of a dynamic system is based on the Basin Entropy concept, developed by Daza et al. \cite{daza2016Basin}, and can be applied here in a complementary way to the verification of the Wada property. In this approach, a measure to quantify the uncertainty is introduced that provides a sufficient condition for the existence of fractal basin boundaries.
As we can see in FIG 7, the disorder increases as $\Omega$, the angle of aperture decreases. Besides, the basin entropy increases as the scale parameter $\epsilon$ increases (for a fixed value of $\Omega$). Even so, a convergence of the basin entropy is expected when the number of trajectories increases, together of the angle of aperture, $\Omega$, for fixed value of $\epsilon$ \cite{daza2016Basin}.

\begin{figure}
\centering
\resizebox{\columnwidth}{!}%
{\includegraphics{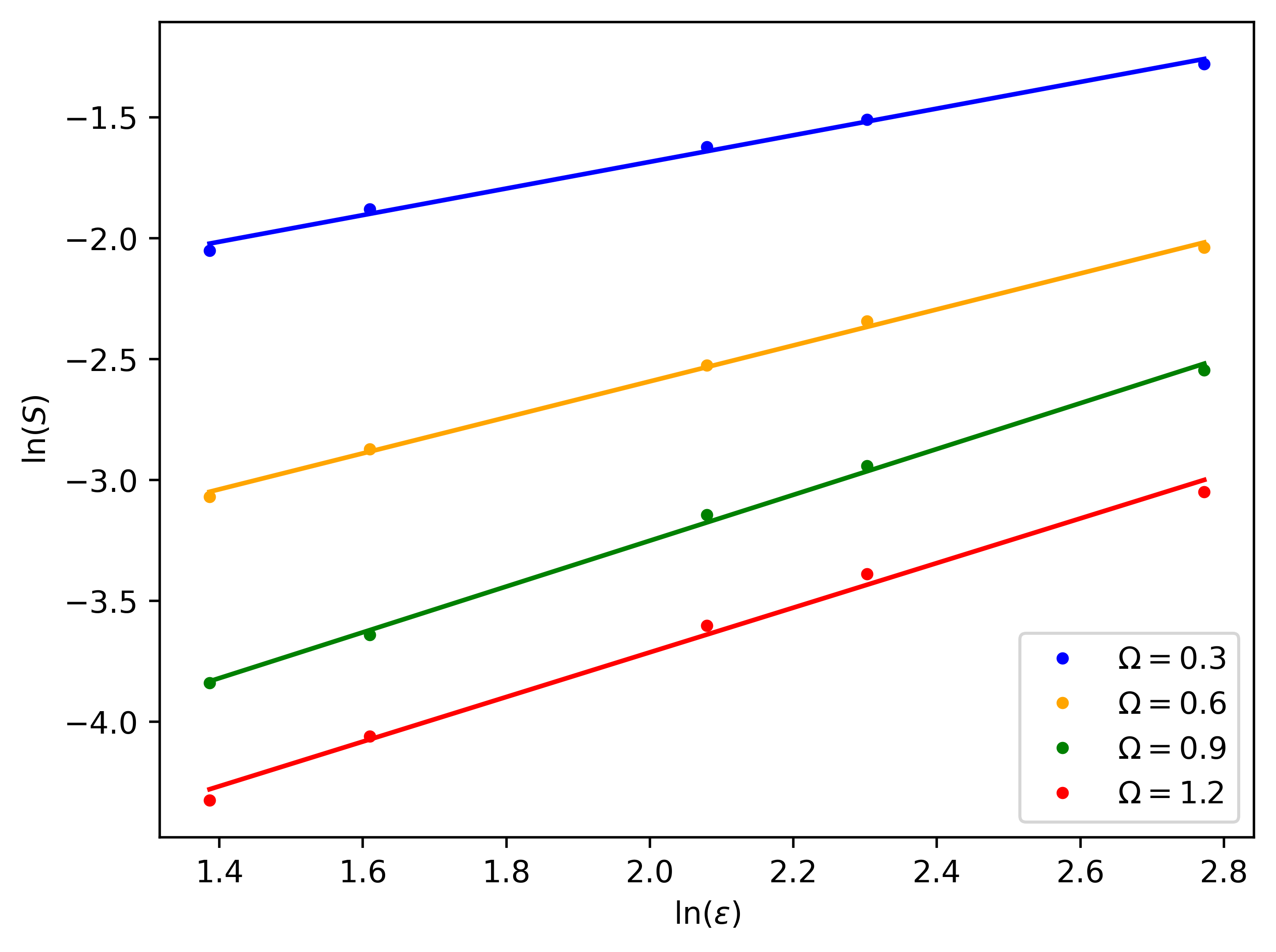}}
 \caption{Basin entropy $S$ given by \eqref{entropia} concerning the set of basins shown in FIG. \ref{fig_basins}. It can be seen that the entropy increases for decreasing $\Omega$. \label{graf_entropia}}
\end{figure}
\begin{figure*}
    \centering
    \includegraphics[height=0.4\textheight]{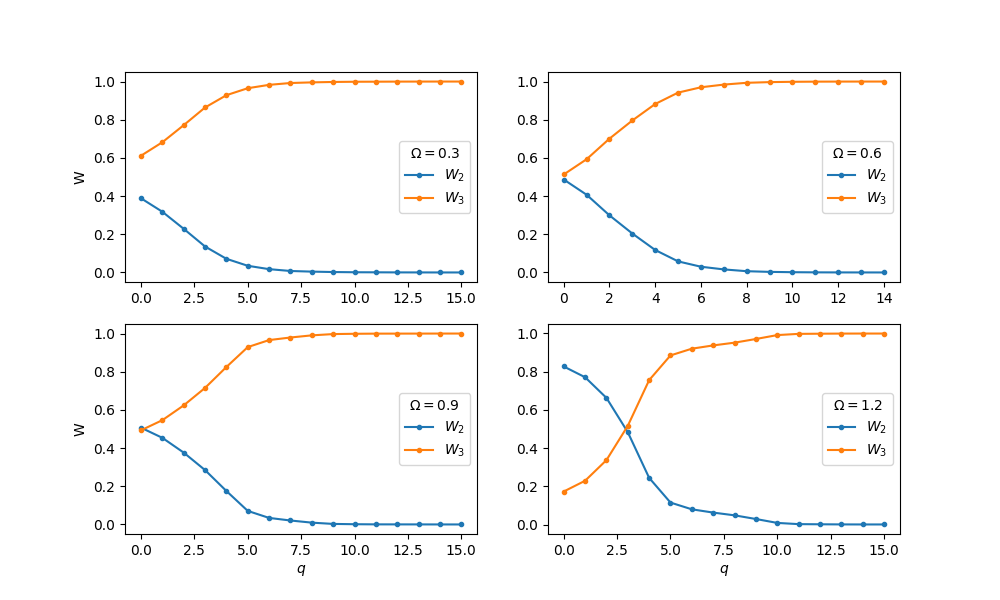}
    \caption{$W_2$ (in blue) and $W_3$ (in orange) convergence of the basins shown in FIG. \ref{fig_basins}. Notice for every case, $W_3$ goes to one.}
    \label{fig_Ws}
\end{figure*}

To obtain this entropy, first the basin must be divided into $N$ square boxes of size $\epsilon\in \mathbb{N}$ all boxes with a number of pixels $\epsilon\times\epsilon$. Then, for each box $j$ consider the probability of an initial condition which exits channel $i$ for a total of $n$ distinct channels
\begin{align*}
&P_{ij}=\frac{\# \mbox{Points with colour i}}{\epsilon_j^2}  & \sum_i^nP_{ij}=1,
\end{align*}
which of course is normalised to $1$. Here $\epsilon_j=\epsilon-n_{0j}$, with $n_{0j}$ the number of trajectories that are not in any of the three channels.

Also, each box has its own entropy defined by
\begin{align*}
    S_j=-\sum_{i=1}^nP_{ij}\ln(P_{ij}),
\end{align*}
for $j=1..N$ since there are $N$ such boxes, while the total entropy is the summation over the entropy of each box
\begin{align}
    S'=\sum_{j=1}^NS_j \label{entropia}
\end{align}
needed to cover the entire basin.

We are interested in the entropy $S'$ relative to the total number of boxes $N$, so we define the basin entropy as
\begin{align*}
S=\frac{S'}{N}
\end{align*}

Again turning to our case, the number of exits is $n=3$ in \eqref{entropia} and FIG. \ref{graf_entropia} shows the basin entropy $S$ given by \eqref{entropia} for each corresponding basin given shown in FIG. \ref{fig_basins}.
\label{sec:basin_entropy}

\section*{Conclusions}
In this article it is obtained the numerical approach to the geodesics in an open hyperbolic manifold with non-trivial topology. We chose as an example the pair of pants together with the leaves, what we call the $\mathbb{Y}$ manifold which is described in section \ref{sec3}.

We have numerically checked that it is in accord with Anosov's famous theorem for geodesic chaos in compact hyperbolic manifolds in the limit that open manifold converges to a closed one with cusps, FIG. \ref{size}. 

To our knowledge this is the first article on numerical geodesics in open manifolds. The standard approach to the understanding of this system is with respect to initial conditions and basin structures. A basin is a set of initial conditions which converge to particular points in phase space, also called channels. The points that reach a particular channel have the same colour and the set with same colour is called the basin attractor to the specif channel. As mentioned in the introduction, it has been verified in several systems that the boundary of these atractors satisfies the Wada property, when the number of colours is bigger than $n>2$. A Wada boundary point satisfies the property that any vicinity of this point contains all other boundaries. 

In this article we specifically checked the Wada property for geodesics in $\mathbb{Y}$ is satisfied by two different techniques, the merging FIG. \ref{merging} and the grid method FIG. \ref{fig_Ws}. We can state that open hyperbolic manifolds have Wada basin boundaries with respect to which exit channel is reached by the geodesic.

Besides checking that the basins are of Wada type, in this article we also obtain the basin entropy for a situation in which the aperture size can be gradually controlled to zero by an angle. We verified that the basin entropy increases with decreasing aperture, which is what can be seen in FIG. \ref{graf_entropia}.

The physical motivation is that $\mathbb{Y}$ could well be a subspace of the $3-D$ spatial section of our Universe for example, $\mathbb{R} \times \mathbb{Y}$ and geodesics the light travel of some distant galaxy. Non-trivial topology in this sense could also contribute to the entropy of the Universe, which according to the most standard theory, is mostly created just after inflation in the reheating process.
\begin{acknowledgments}
The authors would like to thank Nicolas Mocus for his insightful feedback on our codes.
\end{acknowledgments}

\appendix

\section{Euler angles for the corresponding $SU(1,1)$ and $SO(1,2)$\label{A1}}
We give here for reference the parameterization of both groups $SU(1,1)$ acting on $z=x+iy$ through homographies \eqref{homografias}
\begin{align}
&\left( \begin{tabular}{cc}
       $\cosh(\tau/2)$&$\sinh(\tau/2)$  \\
        $\sinh(\tau/2)$ &$\cosh(\tau/2)$
    \end{tabular}\right) 
\left( \begin{tabular}{cc}
       $\cosh(\alpha/2)$& $i$ $\sinh(\alpha/2)$  \\
        -$i$ $\sinh(\alpha/2)$ &$\cosh(\alpha/2)$
    \end{tabular}\right) \nonumber\\
   &\times \left( \begin{tabular}{cc}
        $e^{i\theta/2}$ & $0$ \\
         $0$&$e^{-i\theta/2}$ 
    \end{tabular}\right) 
    \label{A-E}
\end{align}

and its corresponding element in $SO(1,2)$. 

Starting with an infinitesimal transformation on for each parameter in \eqref{A-E} $\delta z$, acting on each \eqref{p_conservado} written as a vector $P=(L,B_1,B_2)$, are locally given by 
\begin{align}
& \delta P=\left( \begin{tabular}{ccc}
       $0$&$0$&$0$  \\
        $0$ &$0$&$-\delta\theta$\\
        $0$ & $\delta\theta$ &$0$
    \end{tabular}\right)P & \delta P=\left( \begin{tabular}{ccc}
       $0$& $\delta\tau$ & $0$  \\
        $\delta\tau$ &$0$&$0$\\
        $0$ & $0$&$0$
    \end{tabular}\right)P \nonumber\\
    &\delta P=\left( \begin{tabular}{ccc}
       $0$& $0$ & $\delta\alpha$  \\
        $0$ &$0$&$0$\\
        $\delta\alpha$ & $0$&$0$
    \end{tabular}\right)P.
\end{align}
It is then well known that any finite transformation is in fact given by the Euler angles of $SO(1,2)$. 
\begin{align}
&\left( \begin{tabular}{ccc}
       $\cosh(\tau)$&$\sinh(\tau)$&$0$  \\
        $\sinh(\tau)$ &$\cosh(\tau)$&$0$\\
        $0$ & $0$&$1$
    \end{tabular}\right) 
\left( \begin{tabular}{ccc}
       $\cosh(\alpha)$&$0$&$\sinh(\alpha)$  \\
       $0$ & $1$& $0$\\
       $\sinh(\alpha)$&$0$ &$\cosh(\alpha)$
    \end{tabular}\right) \nonumber\\
   &\times \left( \begin{tabular}{ccc}
        $1$ & $0$ & $0$ \\
       $0$& $\cos(\theta)$& $-\sin(\theta)$ \\
         $0$&$\sin(\theta)$& $\cos(\theta)$
    \end{tabular}\right) 
\end{align}

\bibliographystyle{apsrev4-2}
\bibliography{refs.bib}

\end{document}